# ARTICLE

# Ferroelectricity Driven by Orbital Resonance of Protons in CH$_3$NH$_3$Cl and CH$_3$NH$_3$Br

Chu Xin Peng,[a][†] Lei Meng,[a][†] Yi Yang Xu,[a][†] Tian Tian Xing,[a] Miao Miao Zhao,[a] Peng Ren,[a][*] Fei Yen[a][*]

The $\beta$ and $\gamma$ phases of methylammonium chloride CH$_3$NH$_3$Cl and methylammonium bromide CH$_3$NH$_3$Br are identified to be ferroelectric via pyroelectric current and dielectric constant measurements. The magnetic susceptibility also exhibits pronounced discontinuities at the Curie temperatures. We attribute the origin of spontaneous polarization to the emergence of two groups of proton orbital magnetic moments from the uncorrelated motion of the CH$_3$ and NH$_3$ groups in the $\beta$ and $\gamma$ phases. The two inequivalent frameworks of intermolecular orbital resonances interact with each other to distort the lattice in a non-centrosymmetric fashion. Our findings indicate that the structural instabilities in molecular frameworks are magnetic in origin as well as provide a new pathway toward uncovering new organic ferroelectrics.

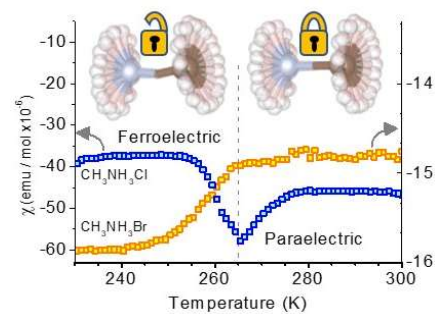

## Introduction

Organic ferroelectrics, let alone multiferroics, remain rare;[1-3] nevertheless, they are desirable as their development should bring forth potential organic devices that are light-weighted, non-toxic, flexible and biodegradable. Ferroelectric behavior may arise from different pathways;[2] one in particular, is when two types of magnetic moments $\mu$ become long-range ordered.[4] The two types of moments interact with each other via antisymmetric exchange interactions which cants their directions breaking spatial-inversion symmetry. Now, these magnetically-driven 'improper ferroelectrics' are all based on the spin configuration of unpaired electrons. Recently, we identified intermolecular magnetic interactions stemming from orbital motion of protons (H$^+$) to play a major role in the structural properties of several hydrogen-bonding materials.[5-8] Orbital magnetic moments $\mu_p$ arise when, for instance, the three protons in each methyl group simultaneously hop to each other's site resembling a quantized rotor.[9] The magnitudes of $\mu_p$ are weak, but they have an orbital resonance aspect and are usually locked to point along a limited number of directions since the proton orbitals are constrained by the crystal field. At threshold temperatures when thermal energy is low enough, intermolecular $\mu_p$–$\mu_p$ interactions occur via resonance so their directions also slightly re-adjust. An effect of this is a structural phase transition with unique distortions that preserve the centrosymmetric character of the lattice but often preclude the structure to be categorized to a particular space group. Now, the ground state of a reorienting CH$_3$NH$_3^+$ ion about the C–N bond situated within a cube of anions (Figure 1a) intrinsically possess two $\mu_p$ with different magnitudes that are coaxial and in tandem to each other (Figure 1b). This is because the C–H (1.10 Å) distances are larger than the N–H (1.04 Å) so the bonding is different and the protons of the CH$_3$ experience a different potential barrier provided the barriers are periodic and near-sinusoidal. Since the NH$_3$ protons experience a different drag, the NH$_3$ reorientations may occur at faster rates so their respective magnetic moments are slightly larger than those of the CH$_3$ ions, namely, $\mu\_{NH3} > \mu\_{CH3}$. At a low enough temperature, when thermal energy is no longer large enough to propel both the CH$_3$ and NH$_3$ groups to reorient more or less as a unit, $\mu\_{CH3}$ and $\mu\_{NH3}$ simultaneously emerge as long-range ordered. The two moments will want to align themselves antiparallel or parallel to each other to minimize energy so their directions will become canted, however, since their magnitudes are different, the degree of canting of the two moments are different which breaks spatial-inversion symmetry. From such, compounds with the CH$_3$NH$_3^+$ ion should become ferrielectric at the least in their ground states.

Arguably one of the simplest solids with the CH$_3$NH$_3^+$ cation are the methylammonium halides which is why we concentrated on CH$_3$NH$_3$Cl (MACl) and CH$_3$NH$_3$Br (MABr). Figure 1b shows the stability regions of the $\alpha$, $\beta$ and $\gamma$ phases for the two compounds.[10,11] The $\alpha$ phase is tetragonal (space group $P4/nmm$) while the lattice structure of the other low temperature phases remain undetermined despite numerous attempts via XRD and neutron scattering experiments.[12,13] Nevertheless, several proton magnetic resonance studies were able to conclude that the CH$_3$ and NH$_3$ groups in the $\beta$ and $\gamma$ phases reorient at different frequencies about the C–N bonds while confined to reorient only along one axis.[14,15] In contrast,

[a.] State Key Laboratory on Tunable Laser Technology, Ministry of Industry and Information Technology Key Laboratory of Micro-Nano Optoelectronic Information System and the School of Science, Harbin Institute of Technology, Shenzhen, University Town, Shenzhen, Guangdong 518055, P. R. China
See DOI: 10.1039/x0xx00000x

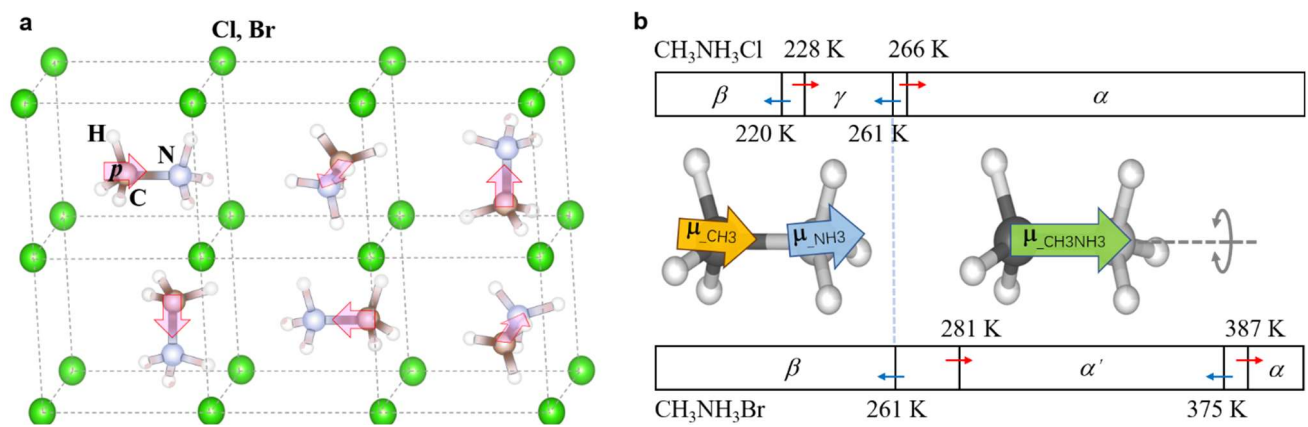

**Figure 1.** (a) Lattice structure of the $\alpha$ phase (*P4/nmm*) of CH$_3$NH$_3$Cl and CH$_3$NH$_3$Br. Each methylammonium cation is surrounded by eight Cl$^-$ ions that are not equidistant to its center of mass so an intrinsic electric dipole moment ***p*** is present. The directions of ***p*** point randomly along the ±*a*, ±*b* and ±*c* directions. (b) Stability regions of the $\alpha$, $\alpha'$, $\gamma$ and $\beta$ phases of CH$_3$NH$_3$Cl and CH$_3$NH$_3$Br. Left and right arrows represent phase transition points during cooling or warming, respectively. Left inset: the ground state of a reorienting methylammonium cation about its C–N bond exhibits two intrinsic magnetic moments, ***μ*_**CH3 and ***μ*_**NH3, because the CH$_3$ and NH$_3$ groups reorient at different rates due to different C–H···Cl and N–H···Cl bonding. Since the directions of ***μ*_**CH3 and ***μ*_**NH3 are not in the least energetic state, a slight canting occurs which breaks spatial-inversion symmetry. Right inset: at higher temperatures when the CH$_3$NH$_3^+$ ion reorients as a whole, there only exists one magnetic moment ***μ*_**CH3NH3 and the centrosymmetric character of the lattice is preserved. In the $\alpha$ and $\alpha'$ phases, the CH$_3$ and NH$_3$ groups reorient coherently, however, in the $\gamma$ and $\beta$ phases, they reorient out of sync so a non-centrosymmetric phase is to be expected.

in the $\alpha$ phase the two groups reorient more or less as a unit while also confined in either the *a*, *b* or *c* axes.[16] We note that a similar trend is encountered in the low temperature phases of the methylammonium lead halides[17] and methylammonium alums.[18] Dielectric constant measurements only exist for the iodine congener CH$_3$NH$_3$I in powdered form but only in the 77–273 K range with a data gap in the 180–220 K range.[19] There are no reports on the magnetic susceptibility at low temperatures because the samples are diamagnetic with no unpaired electrons and proton $\mu_p$–$\mu_p$ interactions were only identified recently to contribute to the magnetization. In the $\beta$ and $\gamma$ phases where thermal energy is no longer large enough to reorient the CH$_3$NH$_3$ ion as a whole, the emergence of the two groups of ***μ*_**CH3 and ***μ*_**NH3 should distort the lattice into a non-centrosymmetric fashion. In this work, we first show that the $\beta$ and $\gamma$ phases of MACl and MABr are ferroelectric via pyroelectric current and dielectric constant measurements. Through measurements of the magnetic susceptibility, we find evidence that the origin of the polarization is driven by proton orbital-orbital interactions. Because the rate of deterioration of the samples are evidently faster in the magnetic susceptibility, the structural deformations are suggested to be magnetically-driven. These results offer new insights as to why many molecular frameworks, for instance, the organic-inorganic perovskite solar cells, suffer structural instabilities.[20] Equally important, our findings provide a new strategy for searching organic and molecular framework ferroelectrics.

## Experimental

Single crystalline plates tabular on [001] were grown from slow evaporation of saturated solutions of CH$_3$NH$_3$Cl (CAS#: 593-51-1) at –18° C and CH$_3$NH$_3$Br (CAS#: 6876-37-5) at 4° C in methanol. The crystals were transparent, irregularly shaped, extremely thin and soft. The polarization was derived from the measured pyroelectric current along the [001] direction of the samples by a Keithley 6517B electrometer. Silver paint contacts were applied onto the two tabular surfaces of the sample to form a parallel plate capacitor-like setup. An electric field of 0.02-1.5 kV/cm was applied in the $\alpha$ phase, then after cooling at 1-3 K/min to the base temperature, the poling field was removed and the contacts shorted for 30 minutes before measuring the current during warming at a rate of 2 K/min. The dielectric constant was extracted from the measured capacitance of the same pair of contacts but with an Agilent E4980A LCR meter at 1 kHz. The temperature environment was controlled by a Physical Properties Measurement System (PPMS) manufactured by Quantum Design, U.S.A. The magnetic susceptibility was measured by the Vibrating Sample Magnetometer (VSM) option of the PPMS. Flakes of around 3–5 mg in weight were attached to a quartz rod with GE varnish with the external magnetic field applied along the *a*-axis.

## Results and discussion

Figure 2a shows the electric polarization $P(T)$ of MACl along the *c*-axis direction with respect to temperature. $P(T)$ was obtained by integrating the measured pyroelectric current $I(T)$ during warming (inset of Figure 2a). A negative poling electric field of $E$ = –0.1 kV/cm was first applied to the crystal during the cooling process which yielded a negative polarization. The process was then repeated but with a positive $E$ of the same magnitude and the result was a polarization with opposite polarity. The reversal of $P(T)$ with an opposite $E$ allow us to classify the $\beta$ and $\gamma$ phases of MACl as ferroelectric. At $T_{C\_\gamma-\alpha}$ = 264 K, when the system was expected to phase transition from $\gamma$ to $\alpha$, $P(T)$ abruptly reduced to near zero indicating that $\alpha$ is paraelectric. The $T_{C\_\gamma-\alpha}$ transition temperature is in excellent agreement with those obtained from heat capacity[10] and differential thermal analysis (DTA).[21] The dielectric breakdown field was near 0.1 kV/cm which is why this value was the highest applied for MACl. Unswitchable polarizations are commonplace in ferroelectric molecular frameworks such as crystals of alum, metal formates and so on.[22-24] A report by Jona *et al*. pointed out that the polarity of ammonium iron alum crystals can nevertheless be switched if the temperature is within 2 K below its transition temperature.[25] We found this to also occur in both MACl and MABr. This is why the base temperature of the MACl measurement shown in Figure 2a was set to 220 K, a few degrees below $T_{C\_\gamma-\beta}$, because in most other attempts when the base temperature was lower, $P(T)$ was unswitchable with an opposite $E$. The largest value recorded for $P(T)$ was 30 μC/cm$^2$ which is in par with other organic molecular ferroelectrics.[26]

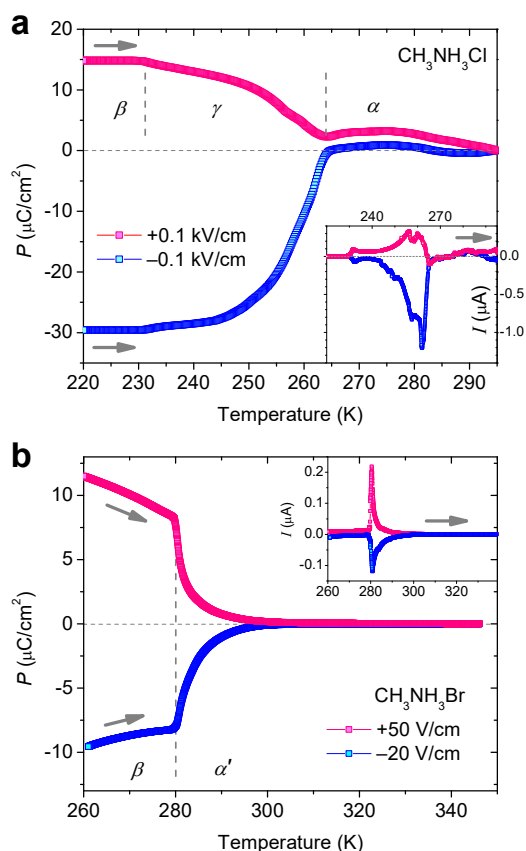

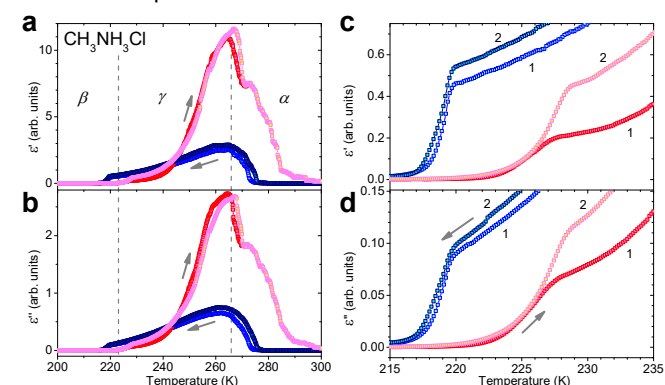

**Figure 2.** Electric polarization $P(T)$ of (a) MACl and (b) MABr along the *c*-axis direction derived from measurement of the pyroelectric current at 2 K/min (insets). The $\gamma$ and $\beta$ phases are identified to be ferroelectric.

Figure 2b shows $P(T)$ obtained for MABr also along the *c*-axis direction from $I(T)$ measurements (inset). A positive $E$ was first applied during cooling prior to measuring $I(T)$ during warming (upper curve in inset of Figure 2b). Immediately afterwards, a negative $E$ was applied and the polarization was nearly completely reversed (lower curves of Figure 2b and inset). These results allow us to conclude that the $\beta$ phase of MABr is also ferroelectric like that of MACl. A clear discontinuity occurred at $T_{C\_\beta-\alpha'}$ = 281 K in $P(T)$ under +$E$ and −$E$ due to the coinciding sharp peaks in $I(T)$ at the same temperature. At $T_{C\_\beta-\alpha'}$, $P(T)$ did not abruptly decrease to zero; it was not until over 300 K that the $\alpha'$ phase became completely paraelectric. Similar to MACl, the base temperature of the MABr measurements in Figure 3b was 260 K, only a few degrees below $T_{C\_\alpha'-\beta}$, because if the crystals were cooled to lower temperatures, the polarization became unswitchable with an $E$ of opposite polarity. As a comparison, the highest recorded polarization was 10 μC/cm$^2$, over 40 times larger than that of Rochelle salt of 250 nC/cm$^2$ near 278 K.[2]

Figures 3a and 3b show the real $\varepsilon'(T)$ and imaginary $\varepsilon''(T)$ parts of the dielectric constant of MACl measured at 1 kHz along the *c*-axis. The transition temperatures $T_{C\_\alpha-\gamma}$ and $T_{C\_\gamma-\beta}$ are clearly evident during cooling and warming in the forms of a maximum and a step-down anomaly in both $\varepsilon'(T)$ and $\varepsilon''(T)$, respectively. Figures 3c and 3d are enlarged areas of Figures 3a and 3b near the $T_{C\_\gamma-\beta}$ region. During cooling the $\gamma$ to $\beta$ transition occurred at 219.6 K while during warming the $\beta$ to $\gamma$ transition at 228.3 K. Above $T_{C\_\gamma-\beta}$, the $\varepsilon'(T)$ and $\varepsilon''(T)$ warming curves never retraced their original paths taken during cooling confirming the metastable nature of the $\gamma$ phase; this feature spills over above $T_{C\_\beta-\alpha}$ which occurred near 261 K and 266 K during cooling and warming, respectively. Peak anomalies over an order of magnitude in $\varepsilon'(T)$ usually indicate an onset of spontaneous polarization with the maximum representing the Curie temperature as observed in Figure 2a. The magnitudes of $\varepsilon'(T)$ and $\varepsilon''(T)$ also progressively change after each cycle similar to the $P(T)$ results if the base temperatures are lower than 220 K. The structural degradation is apparent even to the naked eye as after every cooling cycle, the crystals become bleached because they are fragmented into many small pieces. After several runs, one of the electrodes, the silver paint part, usually detached from the sample.

**Figure 3.** (a) Real $\varepsilon'(T)$ and (b) imaginary $\varepsilon''(T)$ parts of the dielectric constant of MACl at 1 kHz under 1 K/min. (c) and (d) are enlarged regions near the $\gamma$ to $\beta$ phase boundary of $\varepsilon'(T)$ and $\varepsilon''(T)$, respectively. Two successive cooling and warming cycles are displayed labelled by the numbers 1 and 2.

Figures 4a and 4b display $\varepsilon'(T)$ and $\varepsilon''(T)$ of MABr at 1 kHz along the *c*-axis in logarithmic scale, respectively. At the $\alpha'$ to $\beta$ phase boundary, a sharp peak was observed in both $\varepsilon'(T)$ and $\varepsilon''(T)$ at $T_{C\_\alpha'-\beta}$ = 260.5 K; the change in value in $\varepsilon'(T)$ and $\varepsilon''(T)$ were by over 1 and 5 orders of magnitude at the transition, respectively. During warming, similar peaks were observed at $T_{C\_\beta-\alpha'}$ = 280.5 K which coincided to the $I(T)$ and $P(T)$ results. A hysteresis of 20 K and the sharp nature of the peaks (shown in linear scale in the inset of Figure 4a) classify the phase transition as first order. Our recorded $T_{C\_\beta-\alpha'}$ and hysteresis region are in good agreement with those reported for DTA measurements.[21]

Figure 5a shows the magnetic susceptibility $\chi(T)$ as a function of temperature of MACl measured along the *a*-axis under an applied magnetic field of $H$ = 10 kOe. At room temperature the magnitude of $\chi(T)$ was −47 x 10$^{-6}$ emu/mol. Upon cooling, $\chi(T)$ exhibited a minimum of −58 x 10$^{-6}$ emu/mol at $T_{C1}$ = 265 K. At 229 K, $\chi(T)$ suddenly jumped from −40 to −20 x 10$^{-6}$ emu/mol coinciding with the $\gamma$ to $\beta$ phase boundary. Starting from below 220 K, $\chi(T)$ was fairly independent of temperature. Upon warming, the same step but smaller in magnitude was observed at the $\beta$ to $\gamma$ transition point at 230 K. From then onward, $\chi(T)$ during warming for all samples never retraced their cooling paths similar to $\varepsilon'(T)$ and $\varepsilon''(T)$. As reported by Aston et al.,[10] our results indicate that the transition throughout the $\gamma$ phase is quite sluggish. Such discrepancies in $\chi(T)$ indicate the presence of magnetic interactions and magnetic domains at play. Moreover, unlike $\varepsilon'(T)$ and $\varepsilon''(T)$, the second cooling curve of $\chi(T)$ in all measured samples always traced out a completely different path than the first curve. In our view, the canting of $\mu_{\_CH3}$ and $\mu_{\_NH3}$ goes overboard at some of the sites and the lattice cracks initiating the deterioration process of the structure. This suggests that very likely the repeatability crisis and structural instabilities of perovskite solar cells as well as many metal-organic frameworks have magnetic origins.

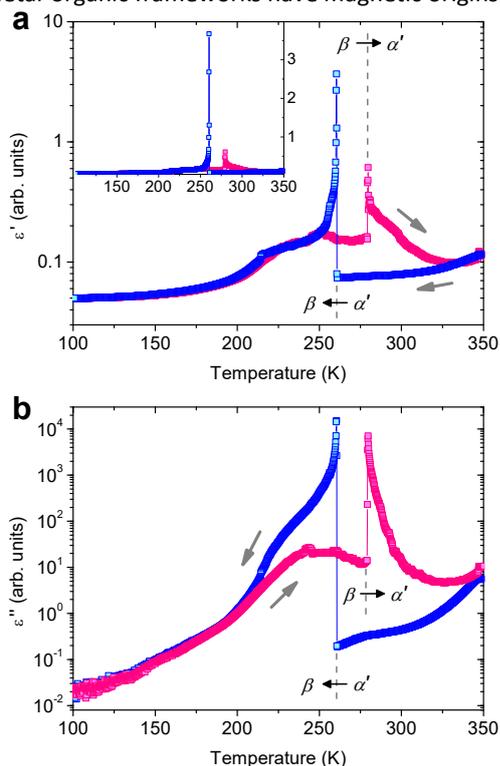

**Figure 4.** (a) $\varepsilon'(T)$ and (b) $\varepsilon''(T)$ of MABr along the *c*-axis direction at 1 kHz under 1 K/min. Vertical dashed lines indicate the transition points between the $\alpha'$ and $\beta$ phases.

Figure 5b shows $\chi(T)$ of MABr also measured along the *a*-axis and under $H$ = 10 kOe. No anomaly was detected at the $\alpha$ and $\alpha'$ phase boundary near 381 K. This is explainable because the transition is believed to only involve a slight distortion of the Br atoms.[11] In contrast, $\chi(T)$ exhibited a step-down feature at $T_{C\_\alpha'-\beta}$ = 258 K at the $\alpha'$ and $\beta$ phase boundary during cooling. The onset of this transition began at 266 K and finished near 250 K. During warming, a cusp developed near $T_{C\_\beta-\alpha'}$ = 281 K quite similar to the $\alpha$ to $\gamma$ transition in MACl during cooling. The critical temperatures obtained from our $\chi(T)$, $P(T)$, $\varepsilon'(T)$ and $\varepsilon''(T)$ measurements all coincide well with each other.

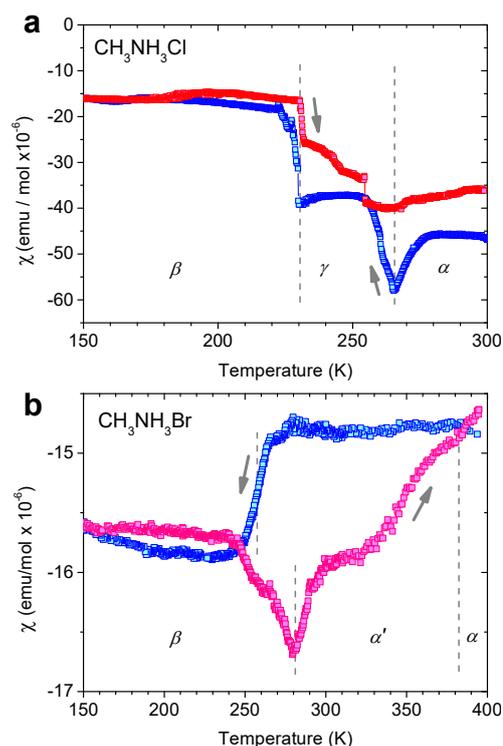

**Figure 5.** Magnetic susceptibility $\chi(T)$ of (a) MACl and (b) MABr under an applied magnetic field of $H$ = 10 kOe with sweeping rates of 1 K/min. All transition temperatures coincide well with those observed in $P(T)$, $\varepsilon'(T)$ and $\varepsilon''(T)$.

The behavior of $\chi(T)$ near the critical points of MACl and MABr cannot be modelled by Landau theory because the magnetization is not the order parameter. The splitting of the magnetic moment of reorienting $CH_3NH_3^+$ into two parts in the $\gamma$ and $\beta$ phases is due to a reduction of random forces in the Langevin equation. From such, the discontinuities in $\chi(T)$ are secondary effects reflecting the dynamics of the lattice. On the other hand, the anomalies in $\chi(T)$ can only be due to intermolecular $\mu_p$–$\mu_p$ interactions for the following reasons: 1) the system possesses no unpaired electrons; 2) if there were unpaired electrons from paramagnetic impurities, then $\chi(T)$ would be positive and behave more or less as $1/T$ during cooling *and* warming; and 3) if there existed no $\mu_p$–$\mu_p$ interactions, namely only Langevin diamagnetism, then $\chi(T)$ should be independent of temperature and at the structural phase transitions, $\chi(T)$ should only decrease by the same percentage of volume contraction. However, the observed changes of 50% in MACl at $T_{C\_\gamma-\beta}$ are unprecedented; even ordinary antiferromagnetic systems rarely exhibit such sharp features upon long-range ordering of the electron spins.

Lastly, all polar ions, most of them organic, should trigger the lattice to become non-centrosymmetric below a critical temperature. However, the rigid framework of the lattice cannot be too soft or too hard, otherwise an orientational glass or fracturing may develop. At the same time, the number of reorienting ions per unit volume must also not be too high, otherwise structural instabilities will ensue such as in the present cases. This explains why replacing a small amount of aspherical $CH_3NH_3^+$ by spherical ions such as $Rb^+$ or $Cs^+$ greatly improved the structural integrity of perovskite solar cells.[27] The tailoring of contemporary designer-molecular frameworks[28-30] should take into account the internal resonances of the lattice.

## Conclusions

To conclude, we presented electric polarization, dielectric constant and magnetic susceptibility measurements as evidence to support the notion that the $\beta$ and $\gamma$ phases in $CH_3NH_3Cl$ and $CH_3NH_3Br$ crystals are ferroelectric and the related spontaneous polarization is magnetically-driven. In a sense, the underlying mechanism may be conceived as a molecular-equivalent of an antisymmetric exchange interaction between the orbital magnetic moments of reorienting $CH_3$ and $NH_3$ groups which slightly cants their equilibrium positions to break spatial-inversion symmetry. Since the two magnetic moments have a tendency to fully align with each other (parallel or antiparallel), parts of the lattice buckle which initiates the structural degradation process. This viewpoint may be employed to explain the plastic nature encountered in many hydrogen-bonding materials as well as exploited to search for ferroelectricity in organic and van der Waals crystals with polar ions.

## Author Contributions

†C. X. Peng, L. Meng and Y. Y. Xu contributed equally.

*Peng Ren (renpeng@hit.edu.cn) and Fei Yen (fyen@hit.edu.cn).

## Conflicts of interest

There are no conflicts to declare.

## Acknowledgements

We acknowledge financial support in part from a General Project grant of the Shenzhen Universities Sustained Support Program grant #GXWD20201230155427003-20200822225417001 and the National Natural Science Foundation of China grant #21702038.

## Notes and references